\documentclass[secnumarabic,amssymb, amsmath,12pt, nofootinbib,tightenlines, nobibnotes, aps, prl]{revtex4}
\usepackage{graphicx}
\usepackage{cancel}
\usepackage{xcolor}
\begin{document}
\newcommand{\be}{\begin{equation}}
\newcommand{\ee}{\end{equation}}
\newcommand{\ba}{\begin{eqnarray}}
\newcommand{\ea}{\end{eqnarray}}
\newcommand{\bea}{\begin{eqnarray*}}
\newcommand{\eea}{\end{eqnarray*}}
\newcommand{\nn}{\nonumber}
\newcommand{\mpi}{m_{\pi}}
\newcommand{\mta}{m_{\tau}}
\newcommand{\tpp}{$\tau^- \to \pi^-\pi^0\nu_{\tau}$}
\newcommand{\eepp}{$e^+e^- \to \pi^+\pi^-$}
\newcommand{\ts}{\textstyle}
\newcommand{\glc}[1]{\textcolor{blue}{#1}}
\newcommand{\nota}[1]{\textcolor{red}{#1}}
\newcommand{\duda}[1]{\textcolor{cyan}{#1}}

\bigskip
\vspace{2cm}
\title{Beyond scalar QED radiative corrections: the $\rho^{\pm}-\rho^0$ width difference, FSR corrections and their impact on $\Delta a_{\mu}^{\rm HVP, LO}[\tau]$}
\vskip 6ex

\author{F.V. Flores-Baez}
\email{francisco.floresbz@uanl.edu.mx}
\affiliation{Facultad de Ciencias Físico Matemáticas, Universidad Autónoma de Nuevo León, Avenida Universidad S/N, Ciudad Universitaria, San Nicolás de los Garza, N.L. CP. 66455, México}

\author{G. López Castro}
\email{gabriel.lopez@cinvestav.mx} 
\affiliation{ {\it Departamento de F\'isica, Centro de Investigaci\'on y de Estudios Avanzados del IPN} \\ {\it Apartado Postal 14-740, 07000 Ciudad de M\'exico, M\'exico}}

\author{Genaro Toledo}
\email{toledo@fisica.unam.mx}
\affiliation{Instituto de Física, Universidad Nacional Autónoma
de México, AP20-364, 01000 Ciudad de México, México}

\bigskip

\bigskip

\begin{abstract}
In a previous paper \cite{Flores-Baez:2007vnd} we have calculated the radiative corrections to $\rho \to \pi\pi$ decays, aiming to estimate the width difference between charged and neutral rho mesons. There, we have used the scalar QED approximation and considered the convection terms to keep the loop contributions finite in the case of charged rho meson decays. Here we compute the radiative corrections by considering the electromagnetic structure of charged mesons and we also include the full Lorentz structure of the electromagnetic vertices. We re-evaluate the width difference of $\rho^{\pm}-\rho^0$ vector mesons and calculate the structure-dependent contributions to Final State Radiation terms in the $e^+e^-\to \pi^+\pi^-$ cross section. Both effects are important inputs for evaluating the isospin breaking corrections in the dominant hadronic vacuum polarization contributions to the muon $g-2$ when using $\tau$ lepton data.
  
 \end{abstract}

\maketitle
\bigskip
\section{Introduction}

In the so-called data-driven approach, the contribution of the hadronic vacuum polarization (HVP) to the muon $g-2$ is largely dominated by the pion electromagnetic form factor below 2 GeV \cite{Bouchiat:1961lbg, Gourdin:1969dm,Aliberti:2025beg}. Information about this form factor can be obtained from $e^+e^-\to \pi^+\pi^-$ annihilation. It was proposed in Ref.~\cite{Alemany:1997tn} that the weak pion form factor measured in $\tau^- \to \pi^-\pi^0\nu_{\tau}$ decay can also be used to evaluate the dominant HVP contribution, provided that isospin breaking (IB) corrections between these two processes are taken into account \cite{Cirigliano:2002pv, Davier:2002dy, Davier:2010fmf}. This is possible because, in the limit of isospin symmetry, the electromagnetic and weak pion form factors coincide exactly for all values of $s$, the pion-pair invariant-mass. Therefore, one can evaluate the HVP contribution to the muon $g-2$ by inserting the normalized $\pi\pi^0$ mass spectrum of $\tau$ decay, $(1/N_{\pi\pi^0})dN_{\pi\pi^0}/ds$, into the dispersion integral, as follows 
\be \label{amutau} 
a_{\mu}^{\rm HVP, LO}[\tau, 2\pi]\! = \!\frac{\alpha^2 m_{\tau}^2}{6|V_{ud}|^2\pi^2} \frac{{\cal B}_{\pi\pi^0}}{{\cal B}_e}\!\! \int_{4m_{\pi}^2}^{m_{\tau}^2} \!\! ds \frac{K(s)}{s} \frac{1}{N_{\pi\pi^0}}\frac{dN_{\pi\pi^0}}{ds}\left(1-\frac{s}{m_{\tau}^2}\right)^{-2}\!\!\!\left( 1+\frac{2s}{m_{\tau}^2}\right)^{-1} \!\!\left[\frac{R_{\rm IB}(s)}{S_{\rm EW}}\right], 
\ee 
 where ${\cal B}_{\pi\pi^0}$ (${\cal B}_e$) is the branching fraction of $\tau$ decay into $\pi\pi^0\nu_{\tau}$ ($e\nu\bar{\nu}$), $S_{\rm EW}$ denotes the short-distance electroweak correction and $K(s)$ is the QED kernel.

The energy-dependent IB corrections to be applied to $\tau$ data are encoded in the last factor of the above equation, and is defined as:
\be \label{rib-factor} 
R_{\rm IB}(s)=\frac{{\rm FSR}(s)}{G_{\rm EM}(s)} \left(\frac{\beta_0(s)}{\beta_+(s)}\right)^3 \left|\frac{F_V(s)}{f_+(s)}\right|^2 \ .
\ee 
The FSR$(s)$ factor contains the final state electromagnetic corrections to the $e^+e^-\to \pi^+\pi^-$ cross section, which is calculated at $O(\alpha)$ in the scalar QED (sQED) approximation; $G_{\rm EM}(s)$ denotes the (model-dependent) long-distance electromagnetic corrections to $\tau^- \to \pi^{-}\pi^0\nu_{\tau}$, currently computed at $O(\alpha)$; $\beta_0(s)$ ($\beta_+(s)$) is the velocity of the pion in the center of mass frame of $\pi^+\pi^-$ in $e^+e^-$ annihilation  ($\pi^-\pi^0$ in the rest frame of $\tau$ decay). Finally, $F_V(s)/f_+(s)$, the ratio of electromagnetic $F_V(s)$ to weak $f_+(s)$ pion form factors, is a correction that accounts for the departure from  isospin symmetry ($F_V(s)/f_+(s)=1$). This ratio depends on the model chosen to describe the pion form factors and, to leading order, on the relevant mass and width difference of $\rho$ mesons, the pion mass difference and the $\rho-\omega$ mixing. 
Current state of the art  evaluations of isospin breaking $\tau \to \pi\pi^0\nu_{\tau}$ decay data can be found in Refs. \cite{Davier:2023fpl,Castro:2024prg,Aliberti:2025beg}.

Isospin  is an approximate but very good symmetry of strong interactions, manifested in the hadron spectrum and their decay properties. IB corrections are usually very small because they are generated, at the fundamental level, by the electromagnetic interactions and the mass difference of $u$ and $d$ quarks. 
Although the pion mass difference is known with high accuracy from experiment, this is not the case for the IB in the $\rho$ meson parameters. According to the most recent White Paper report \cite{Aliberti:2025beg}, current predictions of the HVP contribution to the muon $g-2$ based on tau-data are limited by uncertainties in radiative corrections, due mainly to their scheme-dependence. Other important uncertainties arise from the mass and width differences of $\rho^{\pm}-\rho^0$ mesons entering the ratio of pion form factors, which are poorly known from experiment. The central values and uncertainties associated to the mass and width difference of $\rho$ mesons are dependent of the model used for the form factors in Eq. (\ref{rib-factor}). Thus, using the Gounaris-Sakurai parametrization \cite{Gounaris:1968mw} for these form factors as an example, one estimates  $\pm 0.28\times 10^{-10}$ and $\pm 0.76\times 10^{-10}$ for these uncertainties \cite{Davier:2010fmf}, respectively.

In this work, we focus on the IB in the width of rho-mesons induced by radiative correction effects beyond sQED. Several theoretical calculations and experimental determinations of the difference\footnote{We keep the convention used in Ref. \cite{Davier:2010fmf} for the definition of mass and width differences of neutral and charged rho mesons.} in mass ($\Delta m_{\rho}\equiv m_{\rho^+}-m_{\rho^{0}}$) \cite{Bijnens:1996nq, Feuillat:2000ch, ParticleDataGroup:2024cfk} and width ($\Delta \Gamma_{\rho}\equiv \Gamma_{\rho^0}-\Gamma_{\rho^{\pm}}$) \cite{Flores-Baez:2007vnd, ParticleDataGroup:2024cfk} of neutral and charged rho mesons have been reported in the literature. The data-based determinations of these IB effects, quoted in \cite{ParticleDataGroup:2024cfk}, depend on the specific parameterizations used to model the $2\pi$ invariant mass distributions, and are obtained mainly from $e^+e^-$ annihilations and $\tau$ decay data.  Ref. \cite{ParticleDataGroup:2024cfk} reports  $\Delta \Gamma_{\rho}=(+0.3\pm 1.3)$ MeV and $\Delta m_{\rho}=(+0.7\pm 0.8)$ MeV for the average of measured values. Very recently, from a reanalysis of the most precise $e^+e^-\to \pi^+\pi^-$ and $\tau \to \pi\pi\nu_{\tau}$ data below $\sqrt{s} \leq 0.9$ GeV, Ref. \cite{Davier:2025jiq} obtained $\Delta \Gamma_{\rho} =(-0.58\pm 1.04 )$ MeV and $\Delta m_{\rho}=(+0.30\pm 0.53)$ MeV. In this reanalysis, the normalization of the pion form factors  is left as a free parameter to decorrelate the normalization and the resonance shape in $|F_{\pi}(s)|$, while the shape is modeled with a Gounaris-Sakurai parameterization \cite{Gounaris:1968mw}. The precision obtained in those  data-based determinations of $\Delta \Gamma_{\rho}$ leads to a prediction of the HVP contribution to the muon $g-2$ which compares well in accuracy with lattice calculations \cite{Aliberti:2025beg}. Certainly, improvements in $\Delta \Gamma_{\rho}$ are required to match the precision of experimental $g-2$  \cite{Muong-2:2025xyk}, which is the main motivation of this work. 

Previous theoretical estimates of the mass difference arising from electromagnetic effects have been reported with the results $\Delta m_{\rho} =(-0.7, +0.4)$ MeV \cite{Bijnens:1996nq} and $(0.02\pm 0.02)$ MeV \cite{Feuillat:2000ch}. The theoretical calculation of the width difference arising from the electromagnetic radiative corrections to the $\rho \to \pi\pi$ decays was reported in \cite{Flores-Baez:2007vnd}. Taking into account also the mass difference of the pions in the phase space, it yields $\Delta \Gamma_{\rho}=(+0.76\pm 0.20)$ MeV \cite{Davier:2010fmf}. Radiative corrections were obtained using the sQED  approximation (structureless pion- and rho-photon interactions) and including only the convection-convection terms \cite{Meister:1963zz} in the virtual corrections. In the present work, we extend our previous calculation \cite{Flores-Baez:2007vnd} by including the electromagnetic structure of pions and rho mesons using the vector-meson dominance  (VMD) to model the photon-hadron interactions. We test our calculation of the photon-inclusive radiative corrections to $\rho^0 \to \pi^+\pi^-$ against existing calculations of the one-loop final state radiation (FSR) corrections in $e^+e^- \to \pi^+\pi^-$ \cite{Schwinger:2019zjk, Drees:1990te} for the sQED approximation. As a by-product of our extended calculation, we obtain the structure-dependent effects on the FSR corrections. We point out that the IB contribution of the FSR$(s)$ correction in the evaluation of the HVP using two-pion tau decay data was associated a 10\% uncertainty in Refs. \cite{Castro:2024prg,Davier:2010fmf} due to (missing) structure-dependent virtual electromagnetic corrections.  We also estimate these IB effects in the 2$\pi$ contribution to the leading order HVP using $\tau$ data in the dispersive approach. 

\section{The width difference of rho mesons}

  The total decay width ($\Gamma_{\rho}$) of rho mesons can be defined from Im$\Pi_{\rho}(m_{\rho})=m_{\rho}\Gamma_{\rho}$,  $\Pi(s)$ being the rho meson self-energy \footnote{It is usual to define an off-shell width $m_{\rho}\Gamma_{\rho}(s)={\rm Im}\Pi(s)$. In this case one identifies $\Gamma_{\rho}=\Gamma_{\rho}(m_{\rho})$.}. Thus, it is usually  measured by fitting the resonant lineshape of the $\pi\pi$ invariant-mass distribution using, for instance, the Gounaris-Sakurai parametrization of the pion form factor \cite{Gounaris:1968mw}.  The total width of $\rho$ mesons can also be  defined from the sum over their exclusive final states $f$ as $\Gamma_{\rho}=\sum_f \Gamma(\rho\to f)$, definition that we use in the following.  
  
  Following Ref. \cite{Flores-Baez:2007vnd}, we first estimate the contribution of decay  channels other than $\pi\pi$ and $\pi\pi\gamma$ (we call them ``rest" and include measured $\ell^+\ell^-,\ \pi^0\gamma, \eta\gamma, \pi^0\pi^0\gamma, 3\pi$ and $4\pi$ channels for  $\rho^0$, and $\pi^+\gamma$ for the $\rho^+$ decays). Using the experimental results given in \cite{ParticleDataGroup:2024cfk} we get BR$(\rho^0 \to {\rm rest})=(1.03\pm  0.10)\times 10^{-3}$ and BR$(\rho^+ \to {\rm rest})=(4.5\pm 0.5)\times 10^{-4}$, which leads to a very small contribution to the width difference 
  \ba \label{rest}
  \Delta \Gamma_{\rho}({\rm rest})&=& \Gamma_{\rho^0}\times {\rm BR}(\rho^0 \to {\rm rest})- \Gamma_{\rho^+}\times {\rm BR}(\rho^+ \to {\rm rest}) \nn \\&=&(0.085\pm 0.017)\  {\rm MeV}\ ,
  \ea
  where in the last row we have used the average values of the rho meson total widths  reported in \cite{ParticleDataGroup:2024cfk}. The origin of this difference is basically the $\rho^0\to \eta\gamma$ decay. We will take this difference into account in our final results.

  Therefore, at a level of precision of a few tenths of a percent, the widths of rho mesons are given by their photon-inclusive $\pi\pi$ decay channels, namely
\be
\Gamma_{\rho} \simeq \Gamma[\rho \to \pi\pi(\gamma)]\ ,
\ee
which requires the calculation of the radiative corrections of $O(\alpha)$ to the $\rho \to \pi\pi$ decays. These corrections are calculated in the following sections.
  
The isospin symmetric model for the $\rho\pi\pi$ vertex is given by the interaction Lagrangian ${\cal L}_{\rho\pi\pi} = (g/2)\epsilon_{ijk} \rho^{\mu}_i [(\partial_{\mu}\pi_j)\pi_k- \pi_j(\partial_{\mu}\pi_k)]$, where $i,j,k$ denotes isospin indices, $\epsilon_{ijk}$ is the antisymmetric Levi-Civita tensor and $g$ is the strong coupling constant. If we allow for different values of the $\rho\pi\pi$ coupling $g_{+,0}$ (in the isospin limit $g_+=g_0=g$), the photon-inclusive two-pion decay widths of the rho mesons are given by:
\ba \label{pipiwi1}
\Gamma[\rho^+ \to \pi^+\pi^0(\gamma)]&=& \frac{g_+^2m_{\rho^+}}{48\pi} \beta_+^3(1+\delta_+) ,\\
\Gamma[\rho^0\to \pi^+\pi^-(\gamma)] &=& \frac{g_0^2m_{\rho^0}}{48\pi} \beta_0^3(1+\delta_0), \ \label{pipiwi2} 
\ea
where $\beta_0=\sqrt{1-4m_+^2/m_{\rho^0}^2}$ and $\beta_+=\sqrt{(1-(m_++m_0)^2/m_{\rho^+}^2)(1-\Delta^2/m_{\rho^+}^2)}$ are the pion velocities in the rest frame of decaying $\rho^0$ and $\rho^+$ mesons, respectively. The masses of neutral/charged pion and rho mesons are denoted by $m_{0,+}$ and $m_{\rho^{+,0}}$, respectively, and $\Delta\equiv m_+-m_0$. Finally, $\delta_{0,+}$ contain the radiative corrections of $O(\alpha)$ including real photons  of all allowed energies and depend only on the $\rho$ meson mass and the model used to describe photon-hadron interactions in radiative corrections. 

At first order in the IB parameters, the width difference takes the following form
\be \label{delg}
\Delta \Gamma_{\rho}[\pi\pi(\gamma)] =
\Gamma(\rho^0 \to \pi^+\pi^-) \left[\delta_0-\delta_+ -\frac{2\delta g}{g_0}-\frac{\Delta m_{\rho}}{m_{\rho^0}}\left(\frac{3-2\beta_0^2}{\beta_0^2}\right)-\frac{3}{2}\left(\frac{1-\beta_0^2}{\beta_0^2}\right)\frac{\Delta}{m_+}\right]
\ee
where we have defined
$\delta g \equiv g_+-g_0$ \footnote{Isospin breaking in the strong couplings can arise from both electromagnetic  and $u-d$ quark mass difference corrections to the $\rho\pi\pi$ vertex. In addition to the (dominant) electromagnetic corrections calculated in this work, other model-dependent loop corrections,  involving $a_1$, $\omega$ and the $\rho \pi\gamma$ vertex, can contribute as well. We expect the latter to be negligible due to the small couplings involved in internal vertices. In the present work we do not take into account IB effects in the $\rho\pi\pi$ coupling arising from the $u-d$ mass difference. A reliable estimate at the level of precision required for $\Delta \Gamma_{\rho}$ is beyond the scope of this paper. }. Therefore, the difference in total widths becomes the following
\begin{equation}
\Delta \Gamma_{\rho}= \Delta \Gamma_{\rho}[\pi\pi(\gamma)] +   \Delta \Gamma_{\rho}({\rm rest}) \ .
\end{equation}
The first term of Eq. (\ref{delg}) calculated in this work is going to be used in the last section to evaluate the contribution of the $\Delta \Gamma_{\rho\pi\pi(\gamma)}$ in Table \ref{tabla2} via Eqs. (\ref{amutau},\ref{rib-factor}).

In our previous work \cite{Flores-Baez:2007vnd}, we have computed the radiative corrections $\delta_{+,0}$ using the sQED approximation (point electromagnetic interactions of pions and rho mesons). Furthermore, in the calculation of virtual corrections, only the so-called convection terms \cite{Meister:1963zz} were included for the  electromagnetic vertices of charged particles, which give results that are ultraviolet finite, gauge-invariant and contain all infrared loop divergencies. The calculation of $\delta_0-\delta_+$ reported in Ref. \cite{Flores-Baez:2007vnd} led to $\Delta \Gamma_{\rho}[\pi\pi(\gamma)]\Big|_{\rm \delta_0-\delta_+}=+1.82(20)$ MeV, where a 10\% uncertainty was associated \cite{Davier:2010fmf} to account for missing structure-dependent effects in loop corrections. Therefore, the calculation of the structure-dependent loop corrections will help  reduce that uncertainty. If we add to this result the IB correction due to the pion mass difference (last term within square brackets in Eq. (\ref{delg}), we get $\Delta \Gamma_{\rho}[\pi\pi(\gamma)]= +0.76(20)$ MeV, as it was discussed in \cite{Davier:2010fmf}.

\section{radiative corrections}
The $O(\alpha)$ radiative corrections to $\rho \to \pi\pi$ decays are made up of the virtual (one-loop) corrections and the emission rate of one real photon. These corrections require us to model the photon-hadron interactions at all energies and are, so far, model-dependent. As a first approximation, one can use the vertices by neglecting the structure of hadrons (sQED), as done, for instance, long ago by Schwinger \cite{Schwinger:2019zjk} to compute the corrections to the pion electromagnetic vertex. Owing to current conservation, virtual  corrections are finite in the ultraviolet, although this is not the case for corrections to the weak pion vertex. Similarly, the $O(\alpha)$ corrections to $\rho^0\to  \pi^+\pi^-$ are finite, while the corrections to $\rho^+\to \pi^+\pi^0$ are not \cite{Flores-Baez:2007vnd}. In order to render finite the latter, in Ref. \cite{Flores-Baez:2007vnd} we assumed the so-called convection terms approximation \cite{Meister:1963zz}, where only the radiation off the charge of spin-1 particle \footnote{We use the general form of the electromagnetic vertex for a spin-1 particle as given in Ref. \cite{Kim:1973ee, Hagiwara:1986vm}, $\Gamma_{\alpha\beta\mu}= (p_1+p_2)_{\mu}g_{\alpha\beta}+2(k_{\alpha}g_{\mu\beta}-k_{\beta}g_{\mu\alpha})-p_{1\alpha}g_{\mu\beta}-p_{2\beta}g_{\mu\alpha}$ for the $\rho^+_{\alpha}(p_1) \to \rho^+_{\beta}(p_2)\gamma_{\mu}(k)$ transition, where the last two terms do not contribute for on-shell mesons. In the {\it convection approximation} one keeps only the first term of the electromagnetic vertex. } is considered, so that the virtual corrections to $\rho^+ \to \pi^+\pi^0$ are gauge-invariant, ultraviolet finite and contain all the soft-photon divergencies \cite{Queijeiro:1988vn}.  In the present work, we consider the electromagnetic structure of pions and rho mesons using a vector meson dominance (VMD) to model the photon-hadron interaction. This model gives rise to a finite amplitude for $\rho^+ \to  \pi^+\pi^0$ and allows us to consider the full electromagnetic  vertex for the charged rho meson. In the case of  $\rho^0\to  \pi^+\pi^-$ decays, our sQED calculation can be compared to existing calculations of the FSR corrections \cite{Schwinger:2019zjk,Drees:1990te}, while the extended VMD model considered in this work  allows to compute the structure-dependent effects to FSR. 

\subsection{Virtual corrections} 

\begin{figure}[t]
\vspace{0.0cm}
\begin{tabular}{c}
\includegraphics[width=14cm, angle=0]{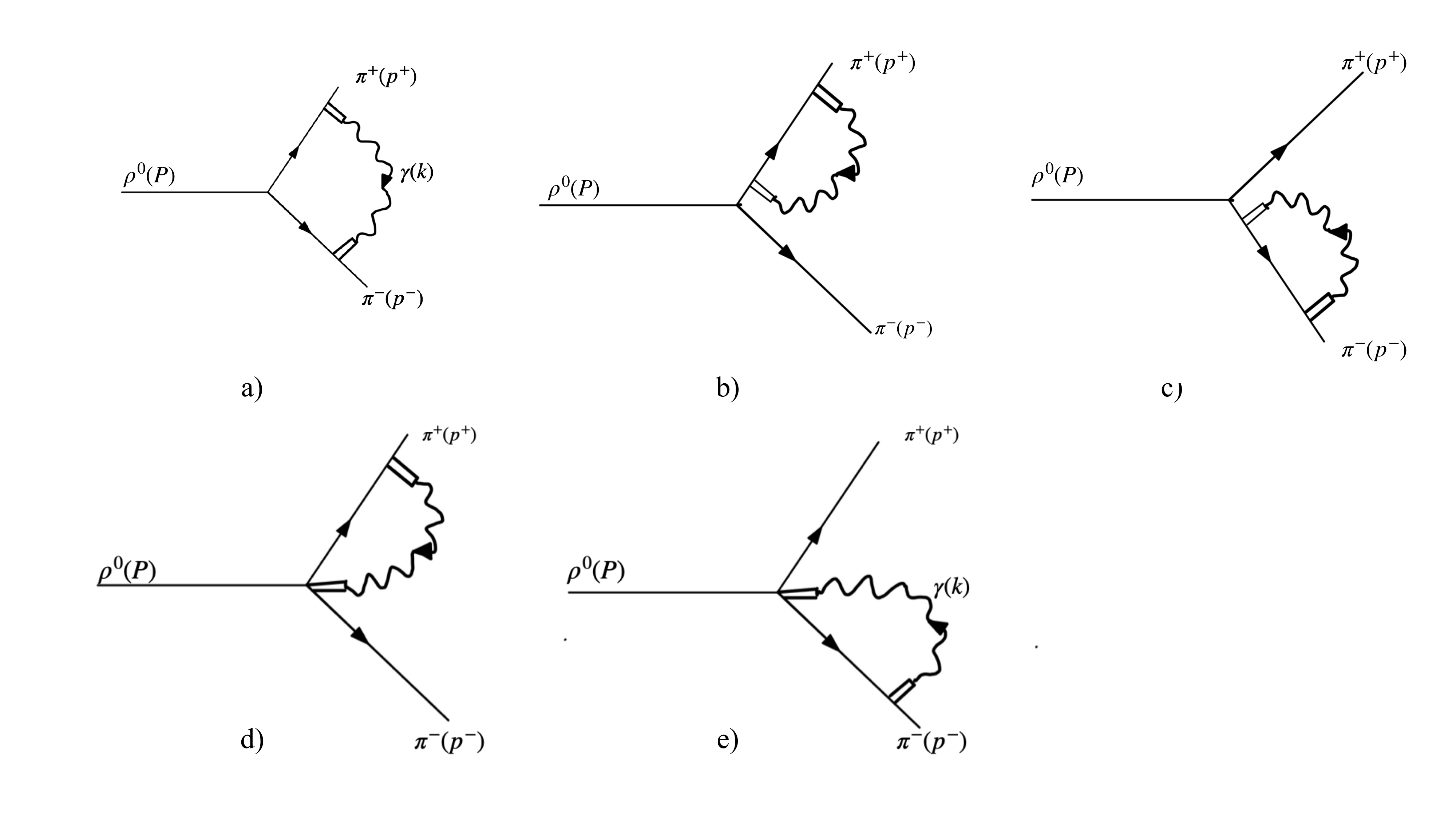} \\
\hline 
\includegraphics[width=14cm, angle=0]
{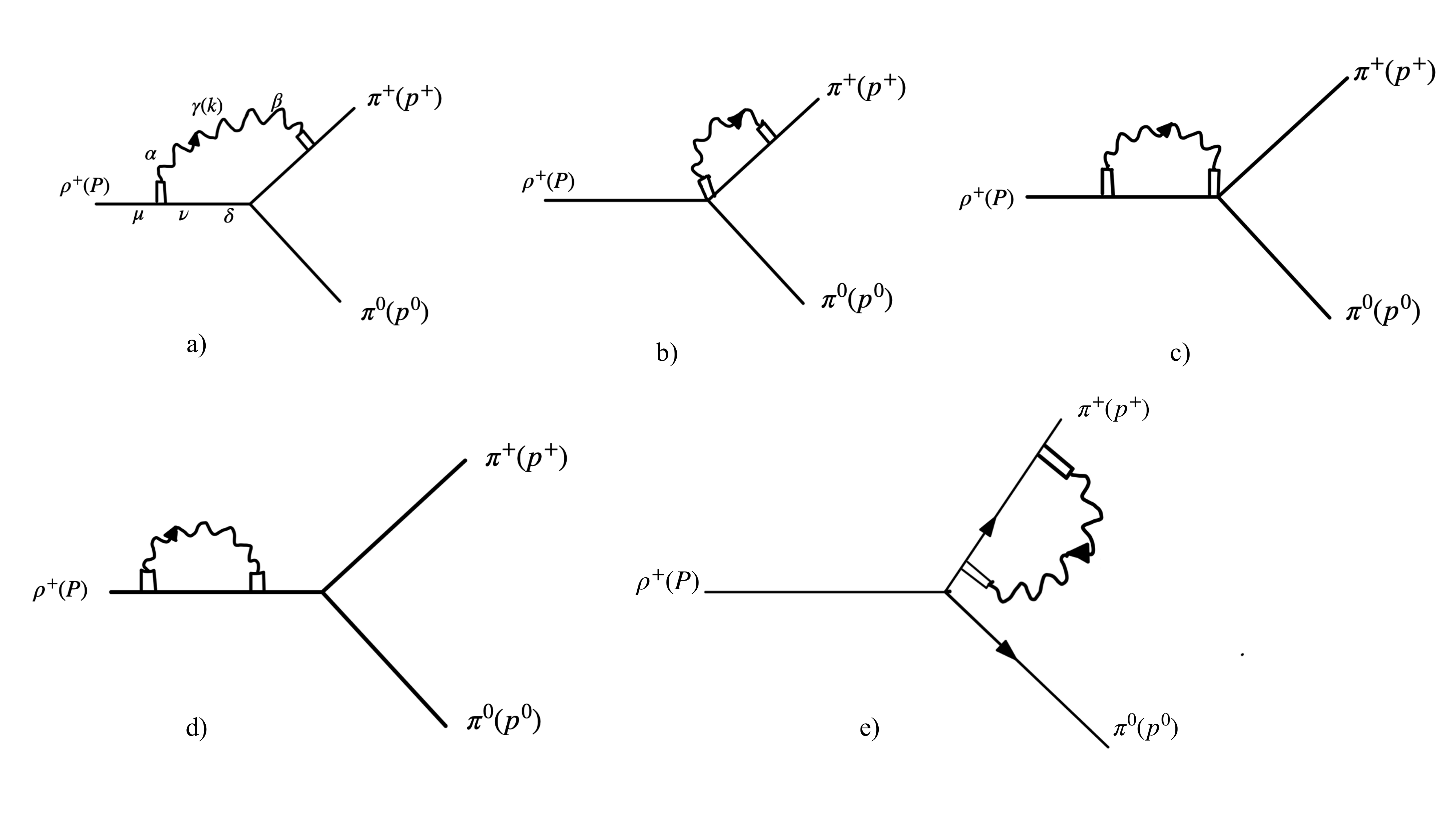} \\
\end{tabular}
\vspace{.5cm}
\caption{Virtual photonic corrections to $\rho^0 \to \pi^+\pi^-$ (upper half) and $\rho^+\to \pi^+\pi^0$ (lower half) decays considering the structure of hadrons in the VMD model. We expect that other model-dependent contributions where virtual mesons can be replaced by $a_1, \omega, \pi$ or $\eta$ mesons, are  very suppressed owing to their small electromagnetic couplings \cite{Flores-Baez:2008owd}.  }
\label{virtualcorr}
\end{figure}

Structure-dependent effects in virtual corrections can be introduced by modifying the photon propagator in loops according to 
\be \label{vmdff}
\frac{1}{k^2} \to \frac{1}{k^2} \left[F(k^2)\right]^2\ ,
\ee
 where a factor $F(k^2)\equiv \sum_{V} a_V\widetilde{m}_V^2/(\widetilde{m}_V^2-k^2)$  ($\widetilde{m}_V^2=m_{V}^2-im_{V}\Gamma_{V}$, where $V$ is the vector resonance with appropriate quantum numbers to mediate the photon-hadron couplings) \cite{Ignatov:2022iou} is attached to the photon coupling to each hadron vertex, as shown in Figure \ref{virtualcorr}. By requiring that the hadron form factors be normalized to the electric charge in the zero momentum limit, we have $F(0)=\sum_V a_V=1$. 
 
 There are two reasons to employ the same form factor for $\pi^{\pm}$ and $\rho^{\pm}$ meson electromagnetic vertices within this generalized vector-meson dominance model (GVMD). First,  the $G$-parity quantum numbers of pions and rho mesons, imply that only  vector-isovector resonances can mediate photon-hadron interactions in Figure \ref{virtualcorr}, therefore we take $V=\rho, \rho', \rho''$. Second, using the same vector-dominance structure for all the vertices as in Eq. (\ref{vmdff}), ensures that the gauge independence of the one-loop amplitude for the $\rho^{\pm}$ case --already satisfied in the sQED calculation-- is preserved (see Appendix A). Unfortunately, the lack of precise data for the $\rho^{\pm}$ electromagnetic vertex prevents a determination of the $a_V$ couplings in this case to validate the model as in the charged pion case \cite{Ignatov:2022iou}. In the model used in this paper, the electromagnetic vertices of pion and rho charged mesons behave asymptotically as $1/k^2$. While this short-distance behavior is expected in QCD for the pion, it is  known that the asymptotic behavior of the rho meson behaves as $1/k^4$ (see for instance Ref. \cite{Braguta:2004kx}). To the extent that intermediate energy structure effects dominate loop integrals, as assumed in this paper, the asymptotic behavior of the form factors would not change our results. 
 
 
 As shown in Ref.~\cite{Ignatov:2022iou}, this  GVMD model allows one to explain the measured  forward-backward charge asymmetry in $e^+e^- \to \pi^+\pi^-$  collisions below 2 GeV \cite{CMD-3:2023alj}, if one includes up three rho-like meson resonances in the photon coupling to hadrons, with the  resonance parameters given in \cite{Ignatov:2022iou}.    Because this form factor does not take into account the energy dependence of the meson widths, among other things, the mass and width parameters obtained from the fit depart \cite{Ignatov:2022iou} from their values extracted using other Breit-Wigner parameterizations (compiled in \cite{ParticleDataGroup:2024cfk}). In our numerical evaluation of loop corrections, we will use the masses and widths of $V$ resonances obtained in Ref. \cite{Ignatov:2022iou}.

 The additional factor $[F(k^2)]^2$ in the amplitudes helps to improve the convergence of loop integrals for hard photons.  The modified photon propagator can be split as
\be \label{sqed+vmd}
\frac{1}{k^2} [F(k^2)]^2
= \frac{1}{k^2}+ \frac{F^2(k^2)-1}{k^2}\ee
The first term on the right-hand side gives rise to the sQED contribution reported in \cite{Flores-Baez:2007vnd} and contains the infrared and ultraviolet divergences, while the second term yields the structure-dependent effects, which are regular at $k^2=0$. Clearly, in the limit $m_V \to \infty$, the structure-dependent  corrections vanish, as it should be
(we have checked this property numerically in our final results) and the results coincide with the sQED calculation. We note that the structure-dependent corrections to  the amplitudes of $\rho^0\to \pi^+\pi^-$, arising from the second term in Eq. (\ref{sqed+vmd}), are ultraviolet finite and do not contribute to the mass-shift of pions or to the renormalization of their wavefunctions.

The decomposition shown in Eq.~(\ref{sqed+vmd}) suggests that the virtual corrections can be split into  sQED and structure-dependent (VMD) contributions according to:
  \begin{equation}
  \delta^{\rm v}_{+,0} = \delta^{\rm v}_{+,0}({\rm sQED})+ \delta^{\rm v}_{+,0}({\rm VMD})\ .\label{split}
  \end{equation}

The sQED virtual corrections of order $\alpha$  in the case of the $\rho^0 \to \pi^+\pi^-$ decay ($\delta^{\rm v}_0({\rm sQED})$), were calculated previously (see Eq. (14) in Ref. \cite{Flores-Baez:2007vnd}). It contains the infrared divergencies, regulated by the photon mass $\lambda$, and is  finite in the ultraviolet (UV) due to current conservation. For future reference, we reproduce it here
\ba \label{virtualn}
\delta_0^{\rm v} ({\rm sQED}) &=& 
\frac{\alpha}{\pi}\left[\pi^{2}\left(\frac{1+\beta^{2}_0}{2\beta_0}\right)-2\left(1+
\ln\left[\frac{\lambda}{m_{\pi^+}}\right]\right)\left(  1+
\frac{1+\beta^2_0}{2\beta_0}\ln\left[\frac{1-\beta_0}{1+\beta_0}\right]\right)\right. \\ 
 &
&\left. -\left(\frac{1+\beta^2_0}{\beta_0}\right)\left[{\rm Li}_2(\beta_0)-
{\rm Li}_2(-\beta_0)\right]-\frac{1+\beta^2_0}{2\beta_0}\left(
{\rm Li}_2\left[  \frac{2}{1+\beta_0}\right]-{\rm
Li}_2\left[\frac{2}{1-\beta_0} \right]\right)\right]\ , \nonumber 
\ea
where  ${\rm Li}_{2}[x]=-\int_{0}^{1} dt\ln[1-xt]/t $ is the dilogarithm function.

The $O(\alpha)$ virtual corrections to $\rho^+ \to \pi^+\pi^0$ were also previously computed, using the convection approximation for the electromagnetic vertices. The result can be found in Eq. (15) of Ref. \cite{Flores-Baez:2007vnd}) where the mass difference $\pi^{\pm}-\pi^0$ was  preserved. A simpler expression  (taken equal masses for pions $m_0=m_+$) that gives a very good approximation is 
\ba \label{dplusv}
\delta_+^{\rm v} ({\rm sQED}) &=& 
\frac{\alpha}{\pi} \left[-1-2\ln\left( \frac{\lambda}{m_{\rho^+}}\right) \left[ 1+\frac{1}{2{\rm v}_+} \ln \left( \frac{1-{\rm v}_{+}}{1+{\rm v}_{+}}\right)\right] +\frac{3}{4}\ln \left( \frac{1-{\rm v}_{+}^2}{4}\right)\right. \nonumber \\ 
&&\left.  +\frac{1}{1-{\rm v}_{+}^2} \left[\ln \left(\frac{4}{1-{\rm v}_{+}^2}\right) -{\rm v}_{+}\ln \left( \frac{1+{\rm v}_{+}}{1-{\rm v}_{+}}\right)\right] \right. \nonumber \\ && \left. +\frac{1}{2{\rm v}_{+}} \left\{ \ln \left(\frac{1-{\rm v}_{+}}{1+{\rm v}_{+}}\right) \left[ 2\ln({\rm v}_{+})+\frac{3}{4}\ln \left( \frac{1-{\rm v}_{+}}{1+{\rm v}_{+}}\right)\right]-\frac{\pi^2}{3}+\ln^2 \left(\frac{1-{\rm v}_{+}^2}{4}\right) \right.\right. \nonumber \\ 
&& \left. \left. +2{\rm Li}_2\left( \frac{1-{\rm v}_{+}}{1+{\rm v}_{+}}\right)+2{\rm Li}_2\left(- \frac{1-{\rm v}_{+}}{1+{\rm v}_{+}}\right)+2{\rm Li}_2\left( \frac{1+{\rm v}_{+}}{2}\right) \right\}\right] \ ,
\ea
where ${\rm v}_{+}\equiv \sqrt{1-4m_{+}^2/m_{\rho^+}^2}$ ($=\beta_+$ in the limit of equal pion masses). All infrared divergencies in the loop corrections are contained in Eq. (\ref{dplusv}), and it is also finite in the UV due to the truncated $\rho\rho\gamma$ vertex. The insertion of the VMD factor into Eq. (\ref{vmdff}) will allow us to consider the full $\rho^+\rho^+\gamma$ vertex (not only the convection term) because this factor improves the convergence of the loop integrals. Therefore, in the case of the $\rho^+ \to \pi^+\pi^0$ decay, only the sum  sQED+VMD in Eq. (\ref{split}) is meaningful.

  The structure-dependent   contributions to virtual corrections --arising from the second term in Eq. (\ref{sqed+vmd})--   can not be easily expressed in simple analytic form. In Appendix B we provide the expression for $\delta_0^{\rm v}({\rm VMD})$ and the full expression for the $\delta_+^{\rm v}$. We assume equal pion masses in the latter case , which is a good approximation at leading order in IB.

We have checked numerically that $\delta^{\rm v}_{0}({\rm VMD})\to 0$ when $m_V\to \infty$, as should be. Next, we consider the real photon corrections necessary to cancel the infrared divergencies.

\subsection{ Real photon corrections}

 The radiative $\rho \to \pi\pi\gamma$ decays have been studied in the past by several authors for different purposes\cite{Singer:1963ae,Bramon:1993yx,LopezCastro:1997dg,LopezCastro:2001apj}. All these calculations include the bremsstrahlung and direct photon emissions as well as model-dependent terms, giving branching ratios in good agreement among them. 
 
 The amplitude of radiative decays can be expressed in the general form:
\be
{\cal M}^{\rm r}(\rho \to\pi\pi\gamma) = {\cal M}_{\rm Low}+{\cal M}_{\rm m.d.}\ ,
\ee
where the Low's amplitude  has the following expansion in terms of the photon momentum $k$, ${\cal M}_{\rm Low}=A/k+B k^0$; the coefficients  $A,\ B$ are determined from the non-radiative process \cite{Low:1975sv}. Their explicit  expressions for the neutral and charged $\rho$ meson decays are given, respectively, in Eqs. (9) and (10) of Ref. \cite{Flores-Baez:2007vnd}. The model-dependent contributions  are summarized in ${\cal M}_{\rm m.d.}$ and start at $O(k)$ \cite{Flores-Baez:2007vnd, LopezCastro:2001apj}. In the model under consideration, they are also mediated by the exchange of light ($\omega,  \sigma, f_0, a_1)$ resonances. Since model-dependent terms contribute to the radiative decay rate at a negligible level compared to the Low's terms \cite{Flores-Baez:2007vnd}, we will neglect them in the following.

 In Ref. \cite{Flores-Baez:2007vnd}, we  included the decay rates of $\rho \to \pi\pi\gamma$ decays as a part of radiative corrections to $\rho\to \pi\pi$ (denoted as $\rho \to \pi\pi(\gamma)$, the photon-inclusive rate). The soft-photon corrections are taken for photon energies lower than $\omega_0$, $\lambda \leq \omega \leq \omega_0$, with $\omega$ the photon energy. They are denoted by $\Gamma_{0,+}^{\rm soft}(\omega_0)$ and depend upon the photon mass regulator $\lambda$, which will cancel the soft photon divergencies appearing in virtual corrections. Their analytic expressions can be found in Eqs. (12) and (13) of Ref. \cite{Flores-Baez:2007vnd}. 
 
 The real-photon corrections for photon energies larger than $\omega_0$, denoted by $\Gamma^{\rm hard}_{0,+}(\omega_0)$, include the contributions of the $Bk^0$ term in Low's amplitude and ${\cal M}_{\rm m.d.}$ as well as their interferences with the soft-photon divergent amplitude. Since the photon energy scale $\omega_0$ is introduced to separate the decay rate into its infrared and finite parts, the sum $\Gamma_{0,+}^{\rm soft}(\omega_0)+\Gamma^{\rm hard}_{0,+}(\omega_0)$ should be independent of the specific value,  provided $\omega_0$ is small (in our numerical evaluations, we verified that this is the case indeed for several values of $\omega_0$). 

 The real photon corrections are thus defined as:
 \ba
 \delta_{0,+}^{\rm \ r} = \frac{\Gamma_{0,+}^{\rm soft}(\omega_0)+\Gamma^{\rm hard}_{0,+}(\omega_0)}{\Gamma(\rho^{+, 0}\to \pi\pi) }\ .
\ea

\subsection{ Radiative corrections and width difference}

  Adding up the virtual and real photon corrections of $O(\alpha)$ we get the radiative corrections:
  \begin{equation} \label{rc0p}
      \delta_{0,+} = \delta^{\rm v}_{0,+} ({\rm sQED+VMD})+\delta_{\rm 0,+}^{\rm \ r} \ .
  \end{equation}
These radiative corrections are ultraviolet finite and free of the soft-photon divergencies. 

In Table \ref{tabla1}, we show the results of the radiative corrections  $\delta_{+,0}$ to the decay rates defined in Eqs. (\ref{pipiwi1}, \ref{pipiwi2}) for values of the $\rho^{+,0}$ meson masses around their measured values ($772-778$ MeV) in steps of 0.5 MeV. We have used $\omega_0=0.5$ MeV for the separation energy of soft and hard real photon emission. The quoted uncertainties stem from the fitted parameters \cite{Ignatov:2022iou} entering the form factors. In the case of  $\rho^0\to \pi^+\pi^-$ decay, the radiative corrections decrease by around 25\% with respect to the sQED approximation calculation reported in \cite{Flores-Baez:2007vnd}. In the case of charged rho meson decays, the corrections are smaller  and have the opposite sign with respect to the result obtained in the convection approximation used in Ref. \cite{Flores-Baez:2007vnd}.

\begin{table}[t!]
 \begin{tabular}{|c|c|c|c|}
 \hline
 \ $m_{\rho^{+,0}}$ (MeV) \  &\ $\delta_0$ ($\times 10^{-3}$)\  & \ $\delta_+$ ($\times  10^{-3}$)\ &\ $\delta_0-\delta_+$ ($\times  10^{-3}$) \  \\
\hline
772.0     &$6.05(2)$ &  $1.44(3)$ & $4.61(4)$ \\
772.5   &  $6.05(2)$ &  $1.44(3)$    & $4.61(4)$\\
773.0     & $6.05(2)$ &  $1.44(3)$ & $4.61(4)$\\
773.5   &$6.05(2)$ & $1.44(3)$  & $4.61(4)$\\
774.0     &$6.04(2)$&$1.44(3)$   &$4.60(4)$\\
774.5   & $6.04(2)$& $1.44(3)$ & $4.60(4)$\\
775.0     & $6.04(2)$& $1.44(3)$  & $4.60(4)$\\
775.5   & $6.03(2)$ &  $1.44(3)$ & $4.59(4)$\\
776.0     & $6.03(2)$& $1.45(3)$& $4.58(4)$\\
776.5 &   $6.03(2)$ & $1.45(3)$& $4.58(4)$\\
777.0     &$6.03(2)$& $1.45(3)$ &$4.58(4)$\\
777.5   &$6.02(2)$& $1.45(3)$& $4.57(4)$\\
778.0     &$6.02(2)$&  $1.45(3)$& $4.57(4)$\\
\hline
 \end{tabular}
\caption{Radiative corrections to the $\rho^0 \to \pi^+\pi^-(\gamma)$ ($\delta_0$) and $\rho^+\to \pi^+\pi^0(\gamma)$ ($\delta_+$) decay rates by varying the mass close to their measured values \cite{ParticleDataGroup:2024cfk}. Uncertainties quoted within round brackets correspond to parametric errors of the form factor  ($a_V, \widetilde{m}_V$). }\label{tabla1}
 \end{table}

   Using our results for the radiative corrections we can calculate the width difference of $\rho$ mesons using Eq. (\ref{delg}). Taking $\Delta =4.5936(5)$ MeV for the mass difference of pions, $m_{\rho}=775$ MeV, $\delta_0-\delta_+=+4.60(4)\times 10^{-3}$ (from Table \ref{tabla1}) and $\Delta m_{\rho}=(+0.7\pm 0.8)$ MeV \cite{ParticleDataGroup:2024cfk}, we get 
   \ba
   \Delta \Gamma_{\rho}[\pi\pi(\gamma)] = (-0.610\pm 0.220) \ {\rm MeV}\ ,  
   \ea
    where we assume $\Gamma(\rho^0 \to \pi^+\pi^-)=150$ MeV in Eq. (\ref{delg}). This width difference receives approximately $+0.690(6)$ MeV as a result of radiative corrections, $-0.196(220)$ MeV due to the rho mass difference, and $-1.104$ MeV due to the pion mass difference in phase space. If we do not take into account the isospin breaking in the rho mass difference $\Delta m_{\rho}=0$ (and set $\delta g=0$), from  Eq. (\ref{delg}) we get
   $\Delta \Gamma_{\rho}=-0.414(6)$ MeV. This value has the opposite sign
   compared to the value used in previous works \cite{Davier:2010fmf,Castro:2024prg} ($\Delta \Gamma_{\rho}=+0.76(0.20)$ MeV)  to estimate its effect on the HVP contribution using tau data, and this is mainly due to a reduction by a factor $\sim 2.6$ in the size of radiative corrections. 

Finally, if we add the width difference arising from other decay channels, given in Eq. (\ref{rest}), we get:
   \begin{equation} \label{wdift}
       \Delta \Gamma_{\rho} = (-0.525\pm 0.220)\ {\rm MeV}\ ,
   \end{equation}
 where all errors have been added in quadrature. 

The result in Eq. (\ref{wdift}) favors a small and negative width difference $\Delta \Gamma_{\rho}$. It is in good agreement with the central value of the data-based determination reported in Ref. \cite{Davier:2025jiq}, $\Delta \Gamma_{\rho}=(-0.58\pm 1.04)$ MeV,  although it is also consistent, within errors, with the average value quoted in \cite{ParticleDataGroup:2024cfk} $\Delta \Gamma_{\rho}=(+0.3\pm 1.3)$ MeV.

\section{FSR corrections to $e^+e^-  \to \pi^+\pi^-$}

 We can study the behavior of radiative corrections for a varying mass of the decaying  particle in the region of relevance to evaluate the contribution of the HVP to the muon $g-2$ ($2m_{\pi} \leq \sqrt{s} \leq 1.8$ GeV).  

As shown in Ref. \cite{Davier:2010fmf} (see also Ref. \cite{Castro:2024prg}) the final-state radiation effects FSR$(s)$ in $e^+e^- \to \pi^+\pi^-$ is one of the isospin breaking corrections to be applied to $\tau^- \to \pi^-\pi^0\nu_{\tau}$ data in order to predict the dominant HVP contribution to the muon $g-2$. The FSR$(s)$ factor, accounting for $\gamma^*(s) \to \pi^+\pi^-(\gamma)$, was calculated long ago in Refs. \cite{Schwinger:2019zjk, Drees:1990te} in the sQED approximation.  The radiative corrections for the $\rho^0 \to \pi^+\pi^-$ reported in  Ref. \cite{Flores-Baez:2007vnd} should be the same as the FSR$(s)$ corrections of Refs. \cite{Schwinger:2019zjk, Drees:1990te} if one allows for a varying mass of the decaying $\rho^0$, namely, 
\begin{equation}
    {\rm FSR}(s)=1+\delta_0(s),
    \end{equation} where $\delta_0(s)$ is given in Eq. (\ref{rc0p}) by replacing $m_{\rho^0}^2\to s $. The method used in Ref. \cite{Flores-Baez:2007vnd} is different from the one used by Schwinger \cite{Schwinger:2019zjk}, since we introduce an energy scale $\omega_0$ to separate soft and hard photons in the evaluation of real photon corrections. Also, our expressions are written in a different form given that the logarithms and di-logarithms functions depend upon different arguments. 

\begin{figure}[t]
\vspace{.0cm}
\includegraphics[width=18cm, angle=0]{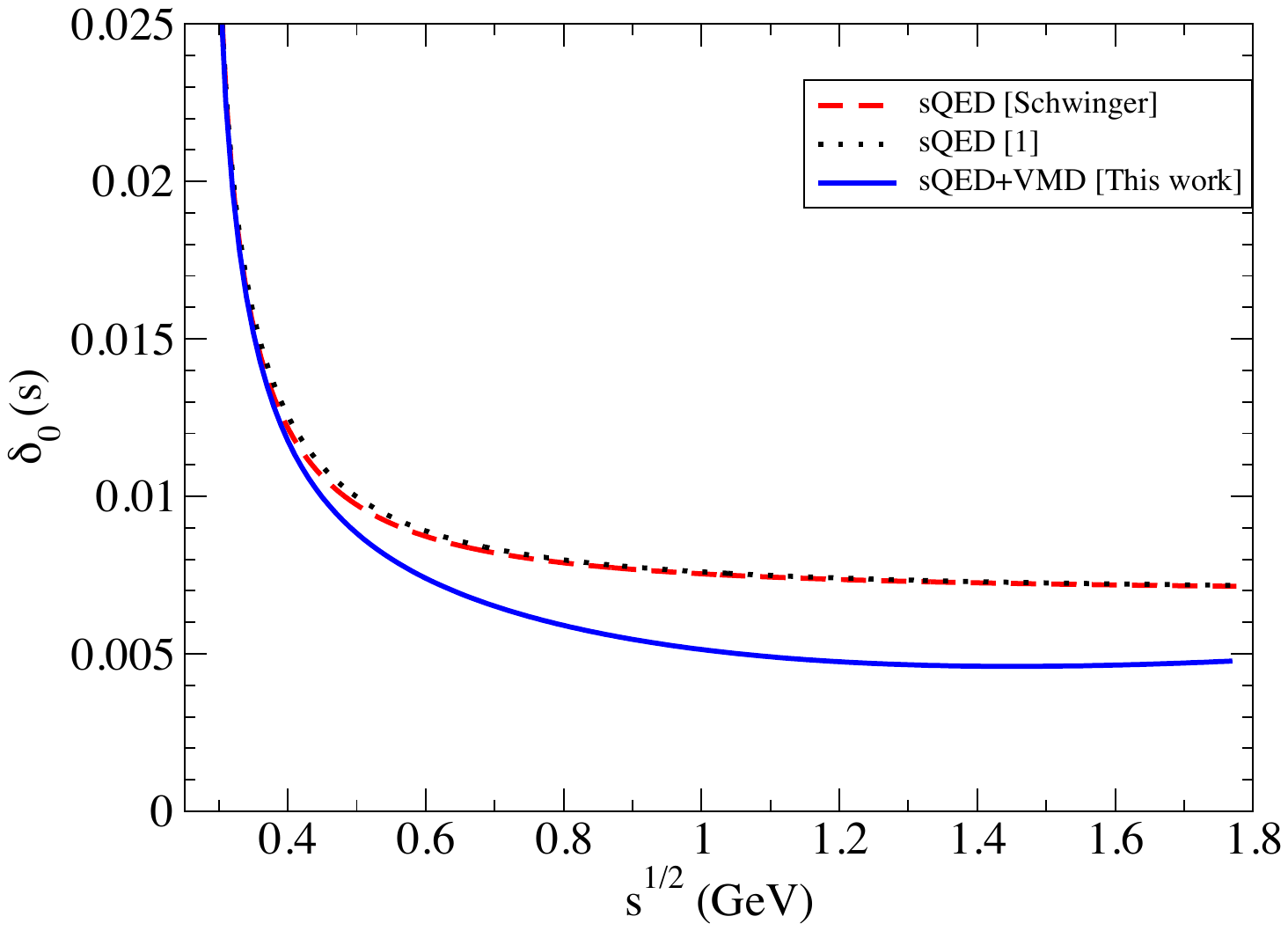}
\vspace{-1.0cm}
\caption{Radiative correction   to $\rho^0\to \pi^+\pi^-(\gamma)$ decay as a function of energy $\sqrt{s}$: scalar QED approximation of Ref.\cite{Flores-Baez:2007vnd} (dotted-line) and the full correction including structure-dependent effects via vector dominance model (VMD) computed in this work (blue solid-line). The FSR$(s)-1$($=\delta_0(s)$) corrections to the $e^+e^-\to \pi^+\pi^-$ cross section computed in Refs. \cite{Schwinger:2019zjk, Drees:1990te} is shown for comparison (red dashed-line).  } 
\label{fsr}
\end{figure}

 In Figure \ref{fsr}, we compare the $\delta_0(s)$ radiative corrections to $\rho^0 \to \pi^+\pi^-$ for a rho meson of mass $2m_{\pi} \leq \sqrt{s}\leq 1.8$ GeV as computed in \cite{Flores-Baez:2007vnd} (dotted-line)  with the FSR$(s)$ corrections of Refs. \cite{Schwinger:2019zjk,Drees:1990te} (red dashed-line). Although both corrections were calculated with a different method and their expressions look different, the FSR corrections clearly give almost identical results\footnote{We use $\omega_0=0.5$ MeV in our results. The agreement is better for smaller values of the photon energy separation scale $\omega_0$. }, thus providing a useful cros check of our calculations in the sQED approximation.
 
 In the same plot (Figure \ref{fsr}) we also display the $\delta_0(s)=$FSR$(s)-1$ correction calculated in the present work (blue solid-line) when the electromagnetic structure of the pions and rho mesons are taken into account via the VMD model.  Structure-dependent effects are not large and contribute around $-30\%$ above $\sqrt{s}=1.2$ GeV.

\begin{figure}[h]
\vspace{.0cm}
\includegraphics[width=18cm, angle=0]{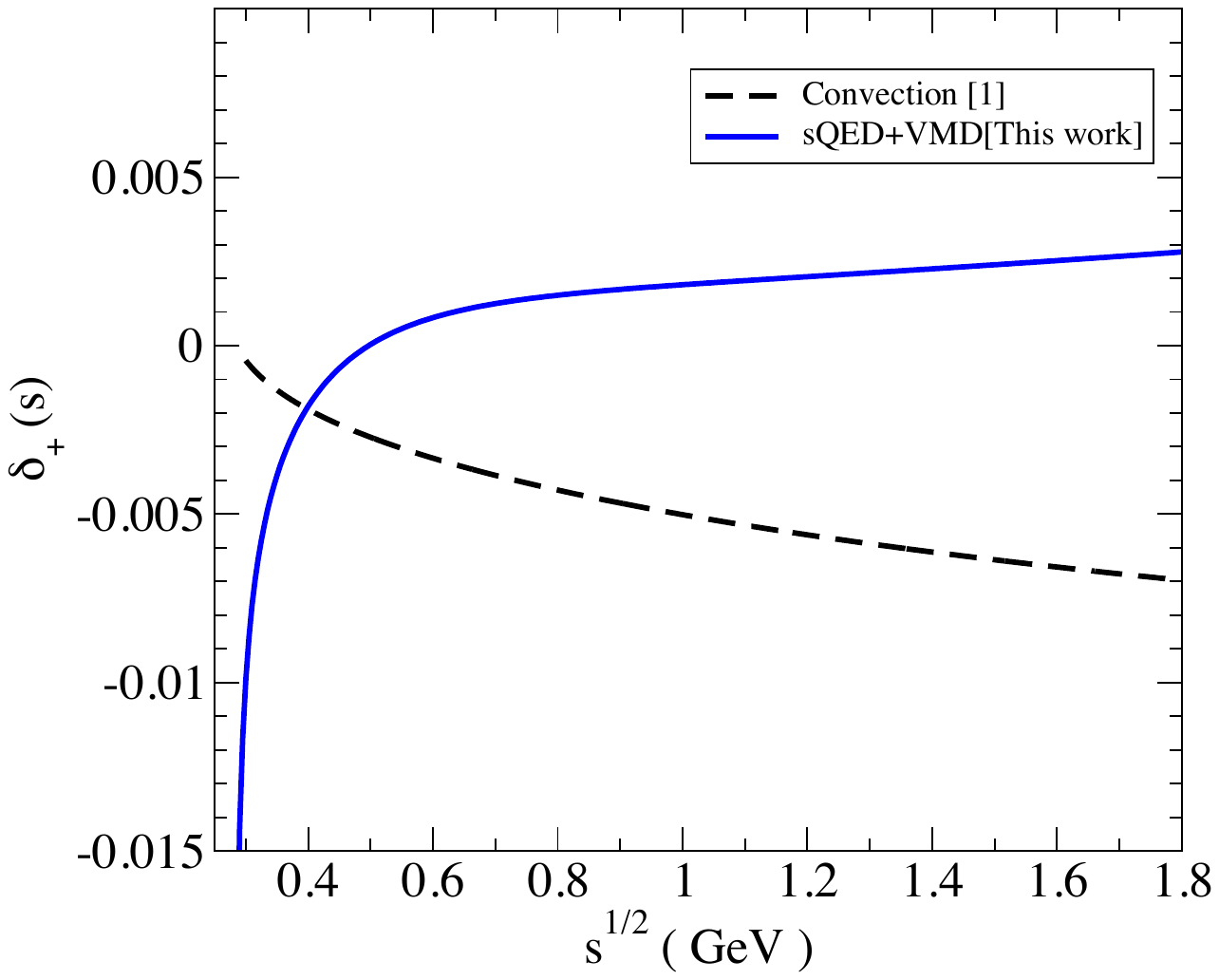}
\vspace{-0.8cm}
\caption{ Radiative corrections to $\rho^+(s)\to \pi^+\pi^0(\gamma)$ decay   computed in this work using VMD (blue solid-line) as a function of energy $\sqrt{s}$. The dashed-line corresponds to the correction in the convection approximation used in Ref. \cite{Flores-Baez:2007vnd}.  } 
\label{delmas}
\end{figure}

Figure \ref{delmas} displays the energy dependence of the radiative correction $\delta_+(s)$ to photon-inclusive $\rho^+\to \pi^+\pi^0(\gamma)$ calculated in this work (blue solid-line). For comparison, the same correction in the convection approximation \cite{Meister:1963zz} used in Ref.~\cite{Flores-Baez:2007vnd} is shown with a dashed-line. Clearly, the use of the full electromagnetic vertex and the electromagnetic structure of charged mesons make the correction  small and positive above $\sqrt{s} \approx 0.5$ GeV.

\section{Impact of FSR and rho width difference on $\Delta a_{\mu}^{\rm HVP, LO} [\tau, 2\pi]$}

It is common to express the shift in the HVP produced by IB corrections to tau data in the dispersive approach, as follows \cite{Aliberti:2025beg, Cirigliano:2002pv, Davier:2002dy,Davier:2010fmf, Davier:2023fpl,Castro:2024prg}
\ba
\Delta a_{\mu}^{\rm HVP, LO}[\tau, 2\pi] &=& \frac{\alpha^2 m_{\tau}^2}{6|V_{ud}|^2\pi^2} \frac{{\cal B}_{\pi\pi^0}}{{\cal B}_e} \int_{4m_{\pi}^2}^{m_{\tau}^2} ds \frac{K(s)}{s} \frac{1}{N_{\pi\pi^0}}\frac{dN_{\pi\pi^0}}{ds}\left(1-\frac{s}{m_{\tau}^2}\right)^{-2}\left( 1+\frac{2s}{m_{\tau}^2}\right)^{-1} \nonumber \\
&&\ \ \ \ \ \  \ \ \ \ \ \ \ \ \ \ \ \ \ \ \ \ \ \ \  \times\left[\frac{R_{\rm IB}(s)}{S_{\rm EW}}-1 \right] \ .
\ea
The different factors entering this equation are explained in the Introduction of this paper and elsewhere \cite{Davier:2010fmf, Aliberti:2025beg}. In the numerical estimates of this work we use ${\cal B}_{\pi\pi^0} =(25.42\pm 0.09)\%$ \cite{Davier:2010fmf}, ${\cal B}_e=(17.84\pm 0.04)\%$ for the branching fractions of the decays of $\tau$ leptons, and $|V_{ud}|=0.97367(32)$ \cite{ParticleDataGroup:2024cfk}. For the spectrum of two-pions in tau decays we use a combined set of ALEPH-CLEO-OPAL-Belle data normalized to the average branching fraction  of the four experiments (${\cal B}_{\pi\pi^0}$ given above) \cite{Davier:2010fmf,private}. 

In this work, we estimate the IB corrections induced by the FSR$(s)$ factor and the so-called $\pi\pi(\gamma)$ decays, including structure-dependent effects. For the latter correction (last factor in Eq. (\ref{rib-factor}))  we use the Gounaris-Sakurai \cite{{Gounaris:1968mw}} parameterization of the pion form factors to compare our results with those of Ref. \cite{Davier:2010fmf}, calculated using the same model. Our results are shown in Table \ref{tabla2}. 

\begin{table}[t!]
 \begin{tabular}{|l|l|l|}
 \hline
 \ \ Source & \ \ FSR [$\times 10^{-10}$]  & \ $\Delta \Gamma_{\rho}(\pi\pi(\gamma))$ [$\times 10^{-10}$]\\
 & & \ \ \ due to $\delta_0-\delta_+$ \\
\hline
\ Schwinger \cite{Schwinger:2019zjk,Drees:1990te} & \ $+4.67(47)$  \cite{Davier:2010fmf} \ & \ $-5.91(59)$ \cite{Davier:2010fmf}\  \\  
\ $\delta_0({\rm sQED})$ \cite{Flores-Baez:2007vnd}  &\  $+4.64(46)$ & \ $-5.97(60)$ \\
\ This work &\  $+3.77(2)$ & \ $-2.22(3)$  \\ 
\hline
 \end{tabular}
\caption{ FSR and $\Delta \Gamma_{\rho}(\pi\pi(\gamma))$ (due to radiative corrections)  contributions to $\Delta a_{\mu}^{\rm HVP, LO}[\tau, 2\pi]$ calculated in this work compared with the sQED approximation of Refs \cite{Schwinger:2019zjk,Drees:1990te} and \cite{Flores-Baez:2007vnd}.}\label{tabla2}
 \end{table} 
 These results for the IB corrections show that the structure-dependent effects are larger than the expected  uncertainties (10\%) associated to them  in Ref. \cite{Davier:2010fmf}. When structure-dependent effects in loop corrections are included, the sum of the IB corrections to $\Delta a_{\mu}^{\rm HVP, LO}[\tau, 2\pi]$, that arise from FSR and the radiative corrections to the $\rho^0 -\rho^{\pm}$ width difference, is shifted by $+2.9\times 10^{-10}$ with respect to previous calculations that use the same form factor model \cite{Davier:2010fmf}. If we add our results for the two corrections calculated in this work (last row Table \ref{tabla2}) to the other IB corrections to be applied to $\tau$ decay data (using the Gounaris-Sakurai parametrization for the IB in pion form factors \cite{Davier:2023fpl}), we get \footnote{The quoted error does not include uncertainties due to (mainly) scheme-dependency in radiative corrections $S_{\rm EW}$, $G_{\rm EM}$ \cite{Aliberti:2025beg} and in the isospin breaking of the $\rho\pi\pi$ couplings due to $u-d$ quark mass difference.  } $\Delta a_{\mu}^{\rm HVP, LO}[\tau, 2\pi]=(-12.0\pm 1.7)\times 10^{-10}$, compared to $(-14.9\pm 1.9)\times 10^{-10}$ as in Ref. \cite{Davier:2023fpl}. 
 
 Taking other IB corrections, updated in Ref.  \cite{Davier:2023fpl}, our final result for the two-pion contribution to the anomalous magnetic moment of the muon based on tau data is:
\begin{equation} \label{amut}
  a_{\mu}^{\rm HVP, LO}[\tau, 2\pi] =(520.2\pm 1.9\pm2.2\pm1.7)\times 10^{-10}\ ,  
 \end{equation}
 which can be compared to $(517.3\pm 1.9\pm 2.2\pm 1.9)\times 10^{-10}$ from Ref. \cite{Davier:2023fpl}. 
 The last uncertainty in both results  stems from the IB corrections. It is slightly reduced by considering the structure-dependent effects in FSR and the width difference, but the central value is sizable modified compared to theoretical uncertainties. Our calculated value shown in Eq. (\ref{amut}) is closer to the result $a_{\mu}^{\rm HVP, LO}[2\pi,$ CMD-3$] =526.0(4.2)\times 10^{-10}$ obtained with measurements of  the CMD-3 collaboration \cite{CMD-3:2023alj,CMD-3:2023rfe} and is larger than  the average of all previous evaluations based on $e^+e^-\to \pi\pi$ cross section data. 

\section{Conclusions  } 

 In this work we have presented  improvements on our previous calculation  \cite{Flores-Baez:2007vnd} of the width difference of $\rho^0-\rho^{\pm}$ mesons induced by radiative corrections. First, while in Ref. \cite{Flores-Baez:2007vnd} we have used a truncated electromagnetic vertex of the charged rho meson to render finite the calculation in the sQED approximation, in the present work we  have included the full expression for the $\rho\rho\gamma$ vertex. Second, we have gone beyond the sQED approximation by including the electromagnetic structure of charged rho and pion mesons. We model this structure via a generalized vector meson dominance \cite{Ignatov:2022iou} model where three  vector mesons ($\rho, \rho',\rho''$) mediate the photon-hadron coupling for virtual photon momenta. We get a finite additional contribution to the sQED radiative corrections for the $\rho^0 \to \pi^+\pi^-$ decays, and also a finite result for radiative corrections in the case of $\rho^{\pm} \to \pi^{\pm}\pi^0$. Structure-dependent effects decrease the radiative correction by 30\% with respect to the sQED correction \cite{Flores-Baez:2007vnd} in $\rho^0 \to \pi^+\pi^-$ decay. The corrections to  $\rho^{\pm} \to \pi^{\pm}\pi^0$ are larger and flips sign with respect to sQED result in the convection approximation \cite{Flores-Baez:2007vnd}, mainly due to the use of a complete $\rho^+\rho^+\gamma$ vertex.

 The width difference of photon-inclusive $\rho \to \pi\pi(\gamma)$  decays induced by radiative corrections is found to be $ \Delta \Gamma_{\rho}[\pi\pi(\gamma)]=+0.69(1)$ MeV (compared to $+1.82(20)$ MeV from Ref. \cite{Flores-Baez:2007vnd}) for $m_{\rho}=775$ MeV. Adding the contributions of isospin breaking due to the pion and rho mesons mass differences and also the contributions of other subleading decay channels we get $\Delta \Gamma_{\rho}=(-0.53\pm 0.22)$ MeV.  This result is in good agreement with the recent
 determination \cite{Davier:2025jiq}, based on two-pion data produced in $\tau$ lepton decay and $e^+e^-$ collisions, and also with the average value quoted in Ref. \cite{ParticleDataGroup:2024cfk} given the large errors of the latter.
 
 Our result for the radiative corrections to the decays of $\rho^0 \to \pi^+\pi^-$ mesons as a function of a varying $\rho^0$ meson mass, namely $1+\delta_0(s)$ was compared to the available calculations (in the sQED approximation) of the FSR corrections to $e^+e^- \to \pi^+\pi$. While the result obtained in the sQED approximation coincides numerically with the well known expressions of Refs. \cite{Schwinger:2019zjk, Drees:1990te}, consideration of the
structure-dependent effects decreases the FSR corrections by about 30\% above the $\rho$ resonance region.

\section{acknowledgments }

G.L.C. acknowledges financial support from SECIHTI project CBF2023-2024-3226. We are thankful to V. Cirigliano, G. Colangelo, M. Davier, M. Hoferichter B. Malaescu, A. Miranda, A. Pich, P. Roig and Z. Zhang for comments and useful discussions.

\section{Appendix A}
The most general  photon propagator reads
\begin{align}
D^{\mu\nu}(k)=\imath\frac{-g^{\mu\nu}+(1-\xi)k^{\mu}k^{\nu}/k^2}{k^2+\imath\epsilon},
\end{align}
where $\xi$ is the gauge parameter (the choice $\xi=1$, used in this work, corresponds to the Feynman gauge). If we assume, for simplicity, the dominance of the $\rho^{\pm}$ and $\pi^{\pm}$  electromagnetic form factors by a single monopolar resonance ($F(k^2)=m_V^2/(m_V^2-k^2)$), the only gauge-dependent pieces of the amplitudes are in 
 the self-energy (SE) and vertex (VX) corrections (Figures \ref{virtualcorr}a,d,e, lower panel). They are given, respectively, by
\begin{align}
\mathcal{M}^{\rm SE}_{\pi}(\xi)&=-(1-\xi)\mathcal{M}_{0}\frac{1}{2}\frac{\alpha}{4\pi
}\left[ \frac{2A_0[m^2_{V}]}{m^2_{V}}-B_0[0,m^2_{V},m^2_{V
}]\right],\\
\mathcal{M}^{\rm SE}_{\rho}(\xi)&=-(1-\xi)\mathcal{M}_{0}\frac{1}{2}\frac{\alpha}{4\pi
}\left[ \frac{2A_0[m^2_{V}]}{m^2_{V}}-B_0[0,m^2_{V},m^2_{V
}]\right],\\
\mathcal{M}^{\rm VX}(\xi)&=(1-\xi)\mathcal{M}_{0}\frac{\alpha}{4\pi
}\left[ \frac{2A_0[m^2_{V}]}{m^2_{V}}-B_0[0,m^2_{V},m^2_{V
}]\right] \ ,
\end{align}
where ${\cal M}_0$ is the tree-level $\rho^{\pm}\to \pi^{\pm}\pi^0$ amplitude. 
Clearly, the gauge-dependence cancels in their sum. This cancellation does not occur if we use different form factors for $\rho$ and $\pi$ mesons. Using, for instance,  monopolar form factors  with different pole masses ($m_b^2 \not =m_a^2$) for $\rho^{\pm}$ and $\pi^{\pm}$ electromagnetic vertices, lead us to the following sum of SE and VX gauge-dependent contributions ($x\equiv (m_b/m_a)^2$)
\begin{equation}
\mathcal{M}(\xi)=(1-\xi)\mathcal{M}_{0}\frac{\alpha}{4\pi}\left( -1-\frac{1+x}{2(1-x)}\ln (x)  \right)\ ,
\end{equation}
which vanishes only for $x=1$ in an arbitrary $\xi$ gauge.

\section{Appendix B}

In this Appendix, we provide the expressions for the virtual one-loop corrections in the GVMD approach \cite{Ignatov:2022iou}. The expressions for the $s$-dependent corrections can be obtained under the replacement $m_{\rho}^2 \to s$.

The virtual one-loop corrections in the case of $\rho^0\to \pi^+\pi^-(\gamma)$ decay are given by 
\begin{align} 
\delta_{0}^{\rm v}({\rm sQED+VMD})&=\delta^{\rm v}_{0}({\rm sQED}) +\frac{\alpha}{\pi}{\rm Re}\left(\sum_{V}a_V^2f_{0}^{\rm VMD}(\widetilde{m}^2_{V},m^2_{\pi^{+}},m^2_{\rho^{0}}) \right. \nonumber \\ &  \hspace*{2.3cm} \left. +\sum_{V< V^{'}}\sum_{V^{'}}a_Va_{V'}f_{0}^{\rm VMD}(\widetilde{m}^2_{V},\widetilde{m}^2_{V^{'}},m^2_{\pi^{+}},m^2_{\rho^{0}})\right)\ , \label{virt1}
\end{align}
where $\delta_{0}^{\rm v}({\rm sQED})$ is given in Eq.~(\ref{virtualn})~\cite{Flores-Baez:2007vnd} and the complex mass parameter is defined as $\widetilde{m}^2_{V}=m^2_{V}-\imath m_V\Gamma_{V}$, with $V,V'=\rho, \rho', \rho''$.  The first term within brackets in Eq.~(\ref{virt1}) is given by
\begin{align}\label{app-vmd}
&f^{\rm VMD}_{0}(\widetilde{m}^2_{V},m^2_{\pi^{+}},m^2_{\rho^{0}})= -\frac{(1-\beta_{0}^2)^2m_{\rho^0}^4}{2(1-{\rm v}^2_{V})^2\beta_{0}^2}D_{0}\Bigl[0,m^2_{\pi^{+}},m^2_{\rho^{0}},m^2_{\pi^{+}},m^2_{\pi^{+}},m^2_{\pi^{+}},\widetilde{m}^2_{V},\widetilde{m}^2_{V},m^2_{\pi^{+}},m^2_{\pi^{+}}\Bigr] \nonumber\\&\left.
\times  \frac{(-2+(1+\beta_{0}^2){\rm v}^2_{V})(-2+\beta_{0}^2(1+{\rm v}^2_{V}))}{(1-\beta^2_{0})(1-{\rm v}^2_{V})}\right.\nonumber\\&\left.
+\frac{m^2_{\rho^{0}}}{\beta^2_{0}}C_{0}\Bigl[m^2_{\pi^{+}},m^2_{\pi^{+}},m^2_{\rho^{0}},m^2_{\pi^{+}},\widetilde{m}^2_{V},m^2_{\pi^{+}}\Bigr]\left(\frac{11-3{\rm v}^2_{V} +\beta^4_{0}(-1+{\rm v}^2_{V})+2\beta^2_{0}(-9+5{\rm v}^2_{V}) }{8(-1+{\rm v}^2_{V})}  \right)\right.\nonumber\\&\left.
+\frac{(1-\beta_{0}^2)m^2_{\rho^0}}{1-{\rm v}^2_{V}}C_{0}\Bigl[0,m^2_{\pi^{+}},m^2_{\pi^{+}},\widetilde{m}^2_{V},\widetilde{m}^2_{V},m^2_{\pi^{+}}\Bigr]\left( \frac{-2+(1+\beta^2_{0}){\rm v}^2_{V}}{\beta^2_{0}(-1+{\rm v}^2_{V})}\right) \right.\nonumber\\&\left.
-\left(B_{0}\Bigl[m^2_{\pi^{+}},m^2_{\pi^{+}},\widetilde{m}^2_{V}\Bigr]-B_{0}\Bigl[m^2_{\rho^{0}},m^2_{\pi^{+}},m^2_{\pi^{+}}\Bigr]\right)\left(\frac{1}{\beta_{0}^2}+1\right)-\frac{m^2_{\rho^0}(1-\beta^2_0)}{2}B_{0}^{'}\Bigl[m^2_{\pi^{+}},m^2_{\pi^{+}},\widetilde{m}^2_{V}\Bigr]\right.\nonumber\\&
-\frac{(1-\beta_{0}^2)m^2_{\rho^{0}}}{2(1-{\rm v}^2_{V})}C_{0}^{'}\Bigl[0,m^2_{\pi^{+}},m^2_{\pi^{+}},\widetilde{m}^2_{V},\widetilde{m}^2_{V},m^2_{\pi^{+}}\Bigr]{\rm v}^2_{V}\ ,
\end{align}
where $\beta_{0}\equiv \sqrt{1-4m^2_{\pi^{+}}/m^2_{\rho^{0}}}$ and  ${\rm v}_{V(V^{'})}\equiv \sqrt{1-4m^2_{\pi^{+}}/\widetilde{m}^2_{V(V^{'})}}$. 

On the other hand, the second term within brackets in Eq.~(\ref{virt1}), symmetric in $(V, V')$, is given by
\begin{align}
f_{0}^{\rm VMD}(\widetilde{m}^2_{V},\widetilde{m}^2_{V^{'}},m^2_{\pi^{+}},m^2_{\rho^{0}})&=f_{0,1}(V,V^{'})+f_{0,1}(V\leftrightarrow V^{'})\nn\\&+f_{0,2}(V,V^{'})+f_{0,2}(V\leftrightarrow V^{'})\nn\\&
+f_{0,3}(V,V^{'})+f_{0,3}(V\leftrightarrow V^{'})\nn\\&+f_{0,4}(V,V^{'})+f_{0,4}(V\leftrightarrow V^{'})+f_{0,5}\ ,
\end{align}
where 
\begin{align}
f_{0,1}(V,V^{'})&=\frac{2(1-{\rm v}^2_{V})}{({\rm v}^2_{V}-{\rm v}^2_{V^{'}})}B_{0}[m^2_{\pi^+},m^2_{\pi^+},\widetilde{m}^2_{V}]\ ,\\
f_{0,2}(V,V^{'})&=-\frac{m^2_{\rho^0}{\rm v}^2_{V}(1-\beta^2_{0})}{({\rm v}^2_{V}-{\rm v}^2_{V^{'}})}B^{'}_{0}[m^2_{\pi^+},m^2_{\pi^+},\widetilde{m}^2_{V}]\ ,\\
f_{0,3}(V,V^{'})&=-\frac{m^2_{\rho^0}}{\beta^2_{0}({\rm v}^2_{V}-{\rm v}^2_{V^{'}})}\frac{(-2+(1+\beta^2_{0}){\rm v}^2_{V})(-2+(1+{\rm v}^2_{V})\beta^2_{0})}{(1-{\rm v}^2_{V})}\nn\\&\times C_{0}[m^2_{\pi^+},m^2_{\pi^+},m^2_{\rho^0},m^2_{\pi^+},\widetilde{m}^2_{V},m^2_{\pi^+}]\ ,\\
f_{0,4}(V,V^{'})&=-\frac{2(-2+\beta^2_{0}+{\rm v}^2_{V})}{\beta^2_{0}({\rm v}^2_{V}-{\rm v}^2_{V^{'}})}B_{0}[m^2_{\pi^{+}},m^2_{\pi^{+}},\widetilde{m}^2_{V}]\ ,\\
f_{0,5}&=\frac{2(1+\beta^2_{0})}{\beta^2_{0}}B_{0}[m^2_{\rho^0},m^2_{\pi^+},m^2_{\pi^+}]\ .
\end{align}
We have checked numerically that when $\widetilde{m}^2_V\to \infty$, the second term in Eq. (\ref{virt1}) vanishes, recovering in this way the sQED approximation result.\\

For the charged meson decay, it is appropriated to separate the loop correction as follows,
\begin{align}
\delta_{+}^{{\rm v}}({\rm sQED +VMD})&=\frac{\alpha}{\pi}{\rm Re}\left( \delta_{\rm point}^{\rm IR-Free}+\delta_{VMD} \right)\ , \label{virt2}
\end{align}
where $\delta_{\rm point}^{\rm IR-Free}$ contains the finite terms that remain after the IR singularity is canceled by the corresponding contribution from  the  soft Bremsstrahlung, and the loop terms that are independent of $\widetilde{m}_{V}$. 
In the isospin limit ( $m_{\pi^{\pm}}=m_{\pi^{0}}$ and $m_{\rho^{\pm}}=m_{\rho^{0}}$), $\beta_{+}\to \overline{\beta}_{+}=\sqrt{1-4m^2_{\pi^+}/m^2_{\rho^+}}$, it takes the following form 
\begin{align}
\delta_{\rm point}^{\rm IR-free}&=-2-\ln \left[\frac{2}{\sqrt{1-\overline{\beta}^2_{+}}} \right]+\frac{1}{2\overline{\beta}_{+}}\ln\left[\frac{1-\overline{\beta}_{+}}{1+\overline{\beta}_{+}} \right]\left(-\ln\left[\frac{2}{\sqrt{1-\overline{\beta}^2_+}} \right]-\frac{1}{2}\ln\left[\frac{1-\overline{\beta}_+}{2(1+\overline{\beta}_{+})} \right]\right.\nonumber\\ &\left.+2\ln\left[ \frac{2\overline{\beta}_{+}}{1+\overline{\beta}_{+}} \right]\right)+\frac{1}{\overline{\beta}_{+}}\left(-\frac{\pi^2}{6}+\rm Li\big[2,\frac{1-\overline{\beta}_{+}}{1+\overline{\beta}_{+}}\big]  +\frac{1}{2}(\ln[\frac{1-\overline{\beta}^2_{+}}{4}])^2 +{\rm Li}\big[2, \frac{1+\overline{\beta}_{+}}{2}\big]\right.\nonumber\\&\left.+{\rm Li}\big[2, \frac{\overline{\beta}_{+}-1}{1+\overline{\beta}_{+}}\big]\right)+\frac{1}{2}B_{0}\Bigl[m^2_{\rho^+},0,m^2_{\rho^+}\Bigr]+\frac{1}{2}B_{0}\Bigl[m^2_{\pi^{+}},0,m^2_{\pi^{+} }\Bigr]\nonumber\\&+\frac{1}{2\overline{\beta}_{+}^2}\left( -B_{0}\Bigl[m^2_{\rho^+},0,m^2_{\rho^+}\Bigr]\left(1-\overline{\beta}_{+}^2\right) +B_{0}\Bigl[m^2_{\pi^{+} },0,m^2_{\pi^{+} }\Bigr]-B_{0}\Bigl[m^2_{\rho^{+} },0,m^2_{\rho^+}\Bigr] \right) \ . \label{uvd1}
\end{align}

The $\delta_{VMD}$ piece in Eq.~(\ref{virt2}) reads
\begin{align}\label{eq:deltafvmd}
\delta_{VMD}&= \sum_{V} a^2_{V}f_{+}^{VMD}(\widetilde{m}^2_{V},m^2_{\pi^{+}},m^2_{\rho^{+}})+\sum_{V<V^{'}} \sum_{V'}a_{V}a_{V^{'}}f^{VMD}_{+}(\widetilde{m}^2_{V},\widetilde{m}^2_{V^{'}},m^2_{\pi^{+}},m^2_{\rho^{+}})\ ,
\end{align}
where
{\allowdisplaybreaks
\begin{align} \label{uvd2}
&f_{+}^{\rm VMD}(\widetilde{m}^2_{V},m^2_{\pi^{+}} ,m^2_{\rho^{+}} )=  m^4_{\rho^+} f_{+,1}D_{0}\Bigl[0,m^2_{\pi^{+}},m^2_{\pi^{+}},m^2_{\rho^{+}},m^2_{\pi^{+}},m^2_{\rho^{+}},\widetilde{m}^2_{V},\widetilde{m}^2_{V},m^2_{\pi^{+}}m^2_{\rho^{+}}\Bigr]\nonumber\\&
+f_{+,2}m^2_{\rho^+}C_{0}\Bigl[0,m^2_{\pi^{+}},m^2_{\pi^{+}},\widetilde{m}^2_{V},\widetilde{m}^2_{V},m^2_{\pi^{+}}\Bigr]
+f_{+,3}m^2_{\rho^+}C_{0}\Bigl[0,m^2_{\rho^{+}},m^2_{\rho^{+}},\widetilde{m}^2_{V},\widetilde{m}^2_{V},m^2_{\rho^{+}}\Bigr]\nonumber\\&
+f_{+,4}m^2_{\rho^+}C_{0}\Bigl[m^2_{\pi^{+}},m^2_{\pi^{+}},m^2_{\rho^{+}},\widetilde{m}^2_{V},m^2_{\pi^{+}},m^2_{\rho^{+}}\Bigr]
+f_{+,5}B_{0}\Bigl[m^2_{\rho^{+}},\widetilde{m}^2_{V},m^2_{\rho^{+}}\Bigr]\nonumber\\&-\frac{1}{2}B_{0}\Bigl[m^2_{\pi^{+}},m^2_{\pi^{+}},\widetilde{m}^2_{V}\Bigr] 
+f_{+,6}\left(B_{0}\Bigl[0,\widetilde{m}^2_{V},\widetilde{m}^2_{V}\Bigr]-B_{0}\Bigl[m^2_{\rho^{+}},\widetilde{m}^2_{V},m^2_{\rho^{+}}\Bigr]\right)\nonumber\\&+\frac{r^3_{+,{\rm v} }}{48} \left(2B_{0}\Bigl[m^2_{\rho^{+}},\widetilde{m}^2_{V},m^2_{\rho^{+}}\Bigr]-\frac{A_{0}[\widetilde{m}^2_{V}]}{r_{+,\rm {v}}m^2_{\rho^+}}-B_{0}\Bigl[0,\widetilde{m}^2_{V},\widetilde{m}^2_{V}\Bigr]\right)\nonumber\\&
-\frac{r^2_{ +,{\rm v} }{\rm v}^2_{V}m^4_{\rho^+}}{4}C_{0}^{'}\Bigl[0,m^2_{\pi^{+}},m^2_{\pi^{+}},\widetilde{m}^2_{V},\widetilde{m}^2_{V},m^2_{\pi^{+}}\Bigr] + \frac{r^2_{+,{\rm v}}}{48}\left(\frac{A_{0}[m^2_{\rho^{+}}]}{m^2_{\rho^{+}}}-B_{0}\Bigl[0,\widetilde{m}^2_{V},\widetilde{m}^2_{V}\Bigr]\right)\nonumber\\& -\frac{m^2_{\rho^+}(1-\overline {\beta}^2_+)}{4}B_{0}^{'}\Bigl[m^2_{\pi^{+}},m^2_{\pi^{+}},\widetilde{m}^2_{V}\Bigr]
-f_{+,7}m^2_{\rho^{+}}B_{0}^{'}\Bigl[m^2_{\rho^{+}},\widetilde{m}^2_{V},m^2_{\rho^{+}}\Bigr]\nonumber\\
&-\frac{5r_{+,{\rm v}}}{12}\left(  \frac{A_0[\widetilde{m}^2_{V}]}{\widetilde{m}^2_{V}}-B_0\Bigl[0,\widetilde{m}^2_{V},\widetilde{m}^2_{V} \Bigr]\right)+\frac{1}{2\overline{\beta}^2_{+}}\left( B_0\Bigl[m^2_{\rho^{+}},\widetilde{m}^2_{V},m^2_{\rho^{+}}\Bigr]-B_0[m^2_{\pi^{+}},m^2_{\pi^{+}},\widetilde{m}^2_{V}]\right)\nonumber\\&
 +\frac{1}{6}r^2_{+,{\rm v}}\left(B_{0}\Bigl[m^2_{\rho^{+}},\widetilde{m}^2_{V},m^2_{\rho^{+}}\Bigr]-B_{0}\Bigl[0,\widetilde{m}^2_{V},\widetilde{m}^2_{V}\Bigr]\right)
 \nn\\&-f_{+,8}m^4_{\rho^+}C_{0}^{'}\Bigl[0,m^2_{\rho^{+}},m^2_{\rho^{+}},\widetilde{m}^2_{V},\widetilde{m}^2_{V},m^2_{\rho^{+}}\Bigr]\ ,
\end{align}
}
where we have defined $r_{+,{\rm v}}\equiv(1-\overline{\beta}_{+}^2)/(1-{\rm v}_{V}^2)$ and
{\allowdisplaybreaks
\begin{align}
f_{+,1}&=\frac{r_{+,{\rm v}}^2}{\rm \overline{\beta}_{+}^2}\left(\frac{(-3+\overline {\beta}^2_{+}+2{\rm v}^2_V)(-1+\overline {\beta}^2_{+}(-1+2{\rm v}^2_V))}{4(1-\overline {\beta}^2_+)}\right)\ ,\\
f_{+,2}&=r_{+,{\rm v}}\left( \frac{3+\overline {\beta}^2_{+}-2(1+\overline {\beta}^2_+){\rm v}^2_V}{4\overline {\beta}^2_{+}(1-{\rm v}^2_V)}\right)\ ,\\
f_{+,3}&=\frac{r^2_{+,{\rm v}}}{48\overline {\beta}^2_{+}(1-{\rm v}^2_V)^2}\left( 24(-1+{\rm v}^2_V)^2(-3+2{\rm v}^2_V)+\overline {\beta}^6_{+}+2\overline {\beta}^4_{+}(-6+5{\rm v}^2_V)\right.\nn\\&\left.+\overline {\beta}^2_{+}(15-18{\rm v}^2_V +4{\rm v}^4_V)\right)\ ,\\
f_{+,4}&=\frac{r_{+,{\rm v}}}{\overline{\beta}^2_{+}} \left(\frac{1+\overline{\beta}^4_{+}+\overline {\beta}^2_{+}(-6+8{\rm v}^2_V -4{\rm v}^4_V)}{4(1-\overline{\beta}^2_+)(1-{\rm v}^2_V)}\right)\ ,\\
f_{+,5}&=\frac{1-\overline{\beta}_{+}^2}{2\overline{\beta}^2_{+}}-\frac{1}{2}\, \\
f_{+,6}&=-\frac{r^2_{+,V}}{24}\, \\
f_{+,7}&=1+\frac{r^3_{+,{\rm v} }}{24} +\frac{r^2_{+, {\rm v} }}{3}\ ,\\
f_{+,8}&=\frac{r^2_{+, {\rm v} }}{48}\left(-68 +r^2_{+,{\rm v} }+16r_{+,{\rm v} }-\frac{48}{r_{+,{\rm v}}} \right)\ .
\end{align}
}

The second term in Eq.~(\ref{eq:deltafvmd}), symmetric in $(V,V')$, has the following expression
\begin{align} \label{uvd3}
f_{+}^{\rm VMD}(\widetilde{m}^2_{V},\widetilde{m}^2_{V^{'}},m^2_{\pi^{+}} ,m^2_{\rho^{+}} )&= 2\left\{ \frac{\textstyle }{\textstyle }  \alpha(V,V^\prime) \left(B_0[m^2_{\pi},m^2_{\pi},\widetilde{m}^2_{V}]+ B_0[m^2_{\rho^{+}},\widetilde{m}^2_{V},m^2_{\rho^{+}}]\right) \right.
\nonumber\\
& \left. - \alpha(V,V^\prime)B_0^{'}[m^2_{\pi},m^2_{\pi},\widetilde{m}^2_{V}]\frac{r_{+,{\rm v}} m^2_{\rho^{+}}{\rm v}^2_{V}}{2(1-\overline{\beta}^2_{+})} \right. \nonumber\\
& \left. +\frac{\alpha(V,V^\prime)({\rm v}^2_{V}-{\rm v}^2_{V^{'}}) }{24}\left[-9\frac{A_0[\widetilde{m}^2_{V}]}{\widetilde{m}^2_{V}}+8B_0[m^2_{\rho},\widetilde{m}^2_{V},m^2_{\rho}]+\frac{A_0[m^2_{\rho}]}{m^2_{\rho}}\right] \right. \nonumber\\
& \left.+\frac{5}{6}\frac{A_0[\widetilde{m}^2_{V}]}{\widetilde{m}^2_{V}}\alpha(V,V^\prime) r_{+,{\rm v}}+\frac{r_{+,{\rm v}}\alpha(V,V^\prime)}{2\overline{\beta}^2_{+}} B_0[m^2_{\pi^{+}},m^2_{\pi^{+}},\widetilde{m}^2_{V}]\right. \nonumber\\
&\left. +\frac{m^2_{\rho^{+}}}{24}B_0^{'}[m^2_{\rho^{+}},\widetilde{m}^2_{V},m^2_{\rho^{+}}]  \alpha(V,V^\prime) \left( 48+64r_{+,{\rm v}}-16r^2_{+,{\rm v}}-r^3_{+,{\rm v}}\right) \right. \nonumber\\
& \left.+\frac{r_{+,{\rm v}}\alpha(V,V^\prime)(1-\overline{\beta}^2_{+}+5)}{6\overline{\beta}^2_{+}(1-\overline{\beta}^2_{+})} B_0[m^2_{\rho^{+}},\widetilde{m}^2_{V^{'}},m^2_{\pi^{+}}] \right. \nonumber\\
& \left.+\frac{r_{+,{\rm v}}\alpha(V,V^\prime)(1-\overline{\beta}^2_{+})}{12}\left(\frac{A_0[\widetilde{m}^2_{V^{'}}]}{\widetilde{m}^2_{V^{'}}} -B_0[m^2_{\rho^{+}},\widetilde{m}^2_{V^{'}},m^2_{\rho^{+}}]\right) \right. \nonumber\\
& \left. +\frac{\alpha(V,V^\prime) r_{+,{\rm v}}}{\overline{\beta}^2_{+}r_{+,{\rm v}^\prime}}\left(B_0[m^2_{\pi^{+}},m^2_{\pi^{+}},\widetilde{m}^2_{V^{'}}]-B_0[m^2_{\rho^{+}},\widetilde{m}^2_{V^{'}},m^2_{\rho^{+}}]\right) \right. \nonumber\\
& \left.+\frac{1-\overline{\beta}^2_{+}}{2\overline{\beta}^2_{+}}\frac{1}{\widetilde{m}^2_{V}-\widetilde{m}^2_{V^{'}}}\widetilde{m}^2_{V}B_0[m^2_{\rho^{+}},\widetilde{m}^2_{V^{'}},m^2_{\rho^{+}}]+ f_{+,9}(V,V^{'}) \right\}\nonumber\\
&+(V \leftrightarrow V^{'})
\end{align}
where the last term $(V \leftrightarrow V^{'})$ symmetrizes the previous expression in curly brackets and we have defined 
\begin{align}
\alpha(V,V^\prime)&=\frac{(1-{\rm v}^2_{V})(1-{\rm v}^2_{V^{'}})}{2({\rm v}^2_{V}-{\rm v}^2_{V^{'}})}\frac{r_{+,{\rm v}^{'}}}{1-\overline{\beta}_{+}^2} \\
f_{+,9}(V,V^{'})&=\alpha(V,V^\prime)m^2_{\rho^{+}}C_{0}\Bigl[m^2_{\pi^{+}},m^2_{\pi^{+}},m^2_{\rho^{+}},\widetilde{m}^2_{V},m^2_{\pi^{+}},m^2_{\rho^{+}}\Bigr]\nn\\& \ \ \ \ \ \ \times\frac{(r_{+,\rm {v}}+2\overline{\beta}^2_{+})(2+r_{+,\rm {v}})}{\overline{\beta}^2_{+}} \ .
\end{align} 
The coefficients of the ultraviolet divergencies in Eqs. (\ref{uvd1}), (\ref{uvd2}) and (\ref{uvd3}) are, respectively, $-(1-3\overline{\beta}_+^2)/2\overline{\beta}_+^2$, $+\sum_V a_V^2(1-3\overline{\beta}_+^2)/2\overline{\beta}_+^2$ and $+2\sum_{V,V'; V < V'}a_Va_{V'}(1-3\overline{\beta}_+^2)/2\overline{\beta}_+^2$. Therefore, it is straightforward to verify that these ultraviolet divergences cancel in the sum of terms in Eq.~(\ref{virt2}) owing to the condition $\sum_V a_V=1$.

The LoopTools package \cite{Hahn:1998yk} was used to  numerically evaluate some of the Passarino-Veltman functions. However, for $C_{0}$ and its derivative $C'_0$ one momentum in their arguments is zero and  they cannot be reliably  evaluated with LoopTools. We followed Ref.~\cite{tHooft:1978jhc} to compute their analytical expressions. The  $C_{0}$-function, which  is free from IR and UV divergences, reads
\begin{align}
C_{0}\Bigl[0,m^2_{\pi^{+}},m^2_{\pi^{+}},\widetilde{m}^2_{V},\widetilde{m}^2_{V},m^2_{\pi^+}\Bigr]&=-\frac{2\left(\ln\left[\frac{4}{1-{\rm v}^2_V}\right]+\frac{1+\rm v^2_{V}}{2{\rm v}_{V}}  \ln\left[\frac{1-\rm v_{V}}{1+\rm v_{V}} \right]\right)}{(1-\overline{\beta}^2_+)m^2_{\rho^{+}}}\ , \\
C_{0}\Bigl[0,m^2_{\rho^{+}},m^2_{\rho^{+}},\widetilde{m}^2_{V},\widetilde{m}^2_{V},m^2_{\rho^+}\Bigr]&=-\frac{2\left(\ln\left[\frac{4}{1-{\rm v}^2_{\rho}}\right]+\frac{1+\rm v^2_{\rho}}{2{\rm v}_{\rho}}  \ln\left[\frac{1-\rm v_{\rho}}{1+\rm v_{\rho}} \right]\right)}{4m^2_{\rho^{+}}}\ ,
\end{align}
where ${\rm v}^2_{\rho}=1-\frac{4m^2_{\rho^{+}}}{\widetilde{m}^2_{V}}$. 

The derivative of $C_{0}$ reads:
\begin{align}
&C_{0}^{'}\Bigl[0,m^2_{\rho^{+}},m^2_{\rho^{+}},\widetilde{m}^2_{V},\widetilde{m}^2_{V},m^2_{\rho^{+}}\Bigr]=-\frac{\partial C_{0}\Bigl[0,p^2,p^2,\widetilde{m}^2_{V},\widetilde{m}^2_{V},m^2_{\rho^{+}}\Bigr]}{\partial p^2}|_{p^2=m^2_{\rho^{+}}}\nonumber\\&=\frac{1}{2m^4_{\rho^{+}}\rm v_{\rho}^{3}}\left[ 2\left(1-\frac{3(1-{\rm v}^2_{\rho})}{2}+\frac{3(1+{\rm v}^2_{\rho})^2}{8}\right) \left( -\rm Arctanh\left[\frac{1}{v_{\rho}} \right]+\rm Arctanh \left[\frac{1+v^2_{\rho}}{2v_{\rho}} \right] \right)\right.\nonumber\\
&\left.+ {\rm v}_{\rho}\left( \frac{1-{\rm v}^2_{\rho}}{2}-\frac{(1-{\rm v}^2_{\rho})^2}{4}+{\rm v}^2_{\rho}\ln\left[\frac{1-{\rm v}^2_{\rho}}{4}\right] \right)\right]\ ,
\end{align}
and $C_{0}^{'}\Bigl[0,m^2_{\pi^{+}},m^2_{\pi^{+}},\widetilde{m}^2_{V},\widetilde{m}^2_{V},m^2_{\pi^{+}}\Bigr]$ is similar to the previous expression under the replacement ${\rm v}_{\rho}\to {\rm v}_V$ in the right-hand-side.

\end{document}